# Title
Anisotropic Strain Relaxation-Induced Directional Ultrafast Carrier Dynamics in $RuO_2$ Films

# Authors
Seung Gyo Jeong,[1,†,*] In Hyeok Choi,[2,†,*] Seungjun Lee,[3,†,*] Jin Young Oh,[4] Sreejith Nair,[1] Jae Hyuck Lee,[5] Changyoung Kim,[5] Ambrose Seo,[6] Woo Seok Choi,[4] Tony Low,[3,*] Jong Seok Lee,[2,*] and Bharat Jalan,[1,*]

# Affiliations
[1]Department of Chemical Engineering and Materials Science, University of Minnesota−Twin Cities, Minneapolis, Minnesota 55455, USA
[2]Department of Physics and Photon Science, Gwangju Institute of Science and Technology (GIST), Gwangju 61005, Republic of Korea
[3]Department of Electrical and Computer Engineering, University of Minnesota, Minneapolis, Minnesota 55455, USA
[4]Department of Physics, Sungkyunkwan University, Suwon 16419, Republic of Korea
[5]Department of Physics and Astronomy, Seoul National University, Seoul 08826, Republic of Korea
[6]Department of Physics and Astronomy, University of Kentucky, Lexington, Kentucky 40506, USA

*Corresponding author. Email: jeong397@umn.edu; jys2316@gm.gist.ac.kr; seunglee@umn.edu; tlow@umn.edu; jsl@gist.ac.kr; bjalan@umn.edu

# Abstract
Ultrafast light-matter interactions inspire potential functionalities in picosecond optoelectronic applications. However, achieving directional carrier dynamics in metals remains challenging due to strong carrier scattering within a multiband environment, typically expected to isotropic carrier relaxation. In this study, we demonstrate epitaxial $RuO_2/TiO_2$ (110) heterostructures grown by hybrid molecular beam epitaxy to engineer polarization-selectivity of ultrafast light-matter interactions via anisotropic strain engineering. Combining spectroscopic ellipsometry, X-ray absorption spectroscopy, and optical pump-probe spectroscopy, we revealed the strong anisotropic transient optoelectronic response of strain-engineered $RuO_2/TiO_2$ (110) heterostructures along both in-plane [001] and [1 1̄ 0] crystallographic directions. Theoretical analysis identifies strain-induced modifications in band nesting as the underlying mechanism for enhanced anisotropic carrier relaxation. These findings establish epitaxial strain engineering as a powerful tool for tuning anisotropic optoelectronic responses in metallic systems, paving the way for next-generation polarization-sensitive ultrafast optoelectronic devices.

# Teaser
Strain-engineered $RuO_2/TiO_2$ heterostructures reveal anisotropic transient optical responses, offering a pathway to next-generation ultrafast optoelectronics.

# MAIN TEXT

# Introduction

The study of light-matter interactions has garnered significant attention in condensed matter physics and materials science, particularly for its role in enabling next-generation optoelectronic technologies. A key area of interest is the engineering of strong polarization selectivity in these interactions, which offers great potential for advancing optoelectronic applications, such as polarization-sensitive photodetectors (*1-4*), linearly polarized light emitters (*5, 6*), and systems for optical communication (*7*), imaging (*8*), and high-speed digital signal transmission (*9*). Central to these applications is the anisotropy of transient photo-response in the picosecond regime, which enables the realization of ultrafast optical switches and modulators. These ultrafast optoelectronic devices, controlled by light polarization, are critical for the next generation of optical information processing technologies (*10, 11*).

Previous studies have largely focused on correlated metallic systems, where anisotropic behavior has been relatively small and restricted to cryogenic temperatures (*12-21*), especially for the transient optical response. Such limitations pose a challenge for practical applications, particularly in room-temperature environments. Recent advancements in the study of low-symmetry insulating two-dimensional (2D) materials have revealed transient optoelectronic anisotropy at room temperature (*10, 22-29*). These present exciting opportunities for developing scalable, stable ultrafast optoelectronic devices capable of functioning under ambient conditions. However, conventional fabrication methods for 2D materials, including mechanical exfoliation (*30, 31*), have encountered major challenges related to scalability and environmental stability, which hinder their practical implementation in large-scale applications. In addition, the anisotropy of materials is inherently determined by the crystal structure, severely limits the degree of tunable anisotropy for engineering specific optical functionalities.

Epitaxial rutile oxide heterostructures provide a promising platform for achieving optoelectronic anisotropy by harnessing anisotropic epitaxial strain. As illustrated in Fig. 1A, $TMO_6$ octahedra (TM: transition metal) in the (110) rutile structure is edge-shared along the [001] direction, where the octahedra at the corner of the unit cell are connected via edge-sharing of equatorial oxygen atoms (gray arrow). In contrast, along the [1 1 0] direction (purple arrow), the octahedra exhibits corner-sharing, involving both equatorial oxygen atoms from the primitive sublattice and apical oxygen atoms from the centered sublattice. This innate structural anisotropy in the rutile structure can lead to the anisotropic electronic band structures resulting in polarization-selective optoelectronic responses. By constructing epitaxial $RuO_2/TiO_2$ (110) heterostructures, additional structural anisotropy can be introduced through epitaxial strain (*32-34*). For instance, the lattice mismatch between $RuO_2$ film grown on $TiO_2$ (110) bulk is $-4.7\%$ along [001], creating compressive strain, and $+2.3\%$ along [1 1 0], generating tensile strain. Furthermore, the large compressive [001] strain can induce the anisotropic strain relaxation starting at 4 nm $RuO_2$ film thickness, while tensile [1 1 0] strain well preserved until 17 nm, leading to anisotropic strain relaxations (*32*). These anisotropic epitaxial strain effects provide a scalable platform for engineering anisotropic light-matter interactions in this metallic system, opening new possibilities for polarization-sensitive ultrafast optoelectronics.

In this paper, we demonstrated both static and dynamic electronic anisotropy engineering in strain-controlled $RuO_2$ films via epitaxial design. Using hybrid molecular beam epitaxy (hMBE), we synthesized atomically flat $RuO_2$ films (*32, 35*) of varying thicknesses on $TiO_2$ (110) (Figs. S1-S3). Combining X-ray absorption spectroscopy (XAS), spectroscopic ellipsometry, and density functional theory (DFT) calculations, we identified that anisotropic strain relaxation effectively modulates the static optical anisotropy in $RuO_2$, starting at ~ 4 nm film thickness, consistent with optical second harmonic generation (SHG) study (*32*). Furthermore, the optical pump-probe spectroscopy reveals significant anisotropy in hot carrier dynamics, demonstrating its tunability via anisotropic strain relaxation. Theoretical analysis on electronic band structures highlighted strain-tunable band nesting in $RuO_2$, as the key mechanism driving directionally selective hot carrier dynamics for anisotropically strain-relaxed $RuO_2$. These findings establish substantial room-temperature optical anisotropy in metallic oxides, offering a new strategy to design polarization-sensitive ultrafast optoelectronic devices using epitaxially strained rutile heterostructures.

**Results**

To quantify the polarization-selective dynamic optical anisotropy, we probe the transient reflectivity change ($\Delta R$) in response to ultrashort laser pulses using a femtosecond laser. By measuring $\Delta R$ along different crystallographic axes (see Fig. 1B), such as the a- and b-axes, the dynamic optical anisotropy between these orientations can be quantified by the parameter $\rho = (\Delta R_a - \Delta R_b)/(\Delta R_a + 2\Delta R_b)$, as similarly defined in previous studies (*27, 36*). This approach further enables the disentangling of the time-dependent contributions of various carrier dynamics (*10, 37*). Within the first picosecond (< 1 ps), carrier scattering dominated by electron-electron scatterings plays a key role, while electron-phonon coupling becomes more prominent in the sub-10 ps regime as carriers thermalize with the lattice (*38*), as schematically shown in the bottom panels of Fig. 1B. These indicate that static electronic and structural anisotropies, as well as their carrier dynamics, are essential for transient optical anisotropy in materials.

We first discuss the crystal structure of $RuO_2/TiO_2$ (110) and the orbital hybridization along two in-plane directions [1 1 0] and [001]. Fig. 1C shows three key orbital states in the (110) plane of rutile $RuO_2$ that influence its electronic structure near the Fermi level (*39-41*). The Ru-4$d$ orbitals split into three $t_{2g}$ states ($d_{x^2-y^2}$, $d_{xz}$, and $d_{yz}$) and two $e_g$ states ($d_{xy}$ and $d_{z^2}$). The $d_{x^2-y^2}$ orbitals, directed toward neighboring Ru atoms, are non-bonding with O-2$p$ orbitals (Left panel of Fig. 1C), whereas the $d_{yz}$ orbitals form $\pi^*$ antibonding with O-2$p_y$ and O-2$p_z$ (Middle panel of Fig. 1C). The coordinate axes defined as, x = [001], y = [1 1 0], and z = [110]. The $d_{xy}$ orbital form $\sigma^*$ antibonding with $sp^2$-type of O orbitals owing to the trigonal-like coordination of Ru and O atoms along the corner sharing direction (Right panel of Fig. 1C). Given these bonding configurations, it is reasonable to expect that anisotropic strain relaxation - which modifies both bond angles and bond lengths - can significantly alter the electronic structure $RuO_2/TiO_2$ (110), leading to tunable anisotropic optical and transport responses.

To investigate this, we performed density functional theory (DFT) calculations. Specifically, we modeled two cases as depicted schematically in Fig. 1D: 1) fully strained $RuO_2$ with lattice constants in both [001] and [1 1 0] directions matching the $TiO_2$ (110) substrate ($a_s = a_{TiO_2}$ and $b_s = b_{TiO_2}$, where $a_s$ and $b_s$ refer to the lattice parameters of a fully strained $RuO_2$ film, and $a_{TiO_2}$ and $b_{TiO_2}$ lattice parameters of the $TiO_2$ (110) substrate along two in-plane directions), and 2) anisotropically strain-relaxed $RuO_2$ (110) with the [001] lattice constant matching bulk $RuO_2$ while [1 1 0] remains fully strained by $TiO_2$ ($a_r = a_{bulk\ RuO_2}$ and $b_r = b_{TiO_2}$, where $a_r$ and $a_{bulk\ RuO_2}$ refer to the relaxed lattice parameter and bulk lattice parameter of $RuO_2$ along [001] direction and $b_r$ is the lattice parameter of $RuO_2$ along [1 1 0]). Fig. 1D illustrates the corner bonding geometry along [1 1 0], showing that $a_r >$ $a_s$ and $\theta_r > \theta_s > 90°$, where $\theta_s$ and $\theta_r$ represent the Ru-O-Ru bonding angles in fully strained and anisotropically strain-relaxed $RuO_2$, respectively. This structural change strongly influences orbital hybridization and electronic properties. We note that the two strain conditions as illustrated in Fig. 1D were chosen to evaluate the effect of anisotropic strain relaxation in $RuO_2/TiO_2$ (110) on electronic structure.

Figures 1E and 1F show the orbital projected density of states for fully-strained and anisotropically strain-relaxed $RuO_2$, as depicted in Fig. 1D, respectively. The solid black lines in the left and right panels represent the total density of states near the Fermi level. While the left panels display projected Ru-4$d$ $t_{2g}$, $e_g$, and O-2$p$ orbitals, the right panels show detailed orbital projections of Ru-4$d$ $t_{2g}$ and $e_g$. The Fermi level of $RuO_2$ is located within the Ru-$t_{2g}$ states, which governs its metallic behavior. More specifically, the projected Ru-4$d$ states in the right panels reveal that the two peaks near the Fermi level are primarily associated with the $d_{x^2-y^2}$ orbital states, and subsequent peaks above the Fermi level are mainly related to $\pi^*$ of $d_{yz}$ and $\sigma^*$ of $d_{xy}$. The $t_{2g}$ states ($d_{x^2-y^2}$ and $d_{yz}$) exhibit a peak-like structure, where the density of states concentrates within a narrow energy range, indicating weak orbital overlap characteristic of lateral π bonding (*39, 40*). In contrast, the $e_g$ state ($d_{xy}$) shows a much broader bandwidth in which the density of states extends across a wide energy range, attributed to the larger head-on overlap of σ-type bonding (*39, 40*). Furthermore, the density of states of the O-2$p$ state exhibits similar behavior to that of the Ru states, indicating strong Ru-O hybridization of $RuO_2$. When strain along [001] is anisotropically relaxed to bulk $RuO_2$ (while strain along [1 1 0] is preserved), the bandwidth of all $\pi^*$ and $\sigma^*$ bonding states become larger and the peak position of $\pi^*$ bonding of $d_{yz}$ states shifts to a higher energy. The former suggests the larger Ru-O orbital overlap in strain-relaxed cases. This further indicates that larger $\theta_r$ by strain-relaxation along the [001] direction, as shown in Fig. 1E, is crucial in determining the Ru-O hybridization rather than the increase in $a_r$. On the other hand, the anisotropic strain-relaxation along [001] also induces the energy lowering of the $d_{x^2-y^2}$ state (indicated by the red arrows in Fig. 1F), consistent with the previous studies (*33, 34, 40*).

To examine the anisotropy of Ru-O orbital hybridization in epitaxial $RuO_2/TiO_2$ (110) heterostructures, we performed XAS as shown in Fig. 2, with total electron-yield (TEY) mode (*42*), at the O *K*-edge since it is strongly coupled with Ru-O hybridized orbitals. Fig. 2A shows the schematic of the setup illustrating how O *K*-edge XAS spectra are acquired with linear horizontal (LH) (electrical field is along [1 1 0] direction) and vertical (LV) (electrical field is along [001]) direction polarized light with an incidence angle of 90°. Before discussing the experimental data, we examine the DFT calculations presented in Fig. 2B. These calculations project the density of states for unoccupied O-2$p$ orbitals along two distinct in-plane crystal orientations: 2$p_x$ along [001] and 2$p_y$ along [1 1 0]. This projection simulates the XAS-related band structures. Since O-2$p_z$ orbitals do not contribute to the O *K*-edge XAS spectrum in the (110) surface normal incidence configuration, our focus is on the 2$p_x$ and 2$p_y$ orbital states. The first peak near 1 eV (in Fig. 2B) is predominantly associated with 2$p_y$ orbital ($\pi^*$-type bonding with $t_{2g}$ states), while the broader peak around 3.7 eV involves a combination of 2$p_x$ and 2$p_y$ orbitals ($\sigma^*$-type bonding with $e_g$ states). Considering the strain effect along the [001] direction between fully strained (top panel) and anisotropically strain-relaxed $RuO_2$ (bottom panel), for both $\pi^*$ and $\sigma^*$ peaks, the bandwidths become broader with strain relaxation, while the peak positions of strain-relaxed $RuO_2$ shift to higher energy compared to that of fully strained $RuO_2$ (vertical dotted lines) indicating increase of the Ru-O hybridization due to anisotropic strain-relaxations, consistent with the conduction band of Ru-4$d$ states in Figs. 1E and 1F.

We now turn to the discussion of experimental results. Fig. 2C shows XAS spectra with LV (upper black) and LH polarization (lower purple) for 2, 4, and 12 nm $RuO_2$ films. Each spectrum was normalized by the peak intensity at 532 eV. The vertical dotted lines for 2 nm $RuO_2$ film (left panel) show the two peak positions at ~529 and ~532 eV of $t_{2g}$-$\pi^*$ and $e_g$-$\sigma^*$ states, respectively, in agreement with the previous rutile studies (40, 43, 44). Especially for LH polarization, [1 1 0] directional electric field shows a much stronger $t_{2g}$-$\pi^*$ peak, consistent with the dominant contribution of $2p_y$ orbital for the $t_{2g}$-$\pi^*$ state shown in Fig. 2B. Interestingly, as $RuO_2$ thickness ($d$) increases, both relative peak intensity ($I(t_{2g})/I(e_g)$) and separation ($\Delta$) between $t_{2g}$-$\pi^*$ and $e_g$-$\sigma^*$ peaks increase (plotted in Figs. 2D and 2E). Furthermore, the thickness dependence of $I(t_{2g})/I(e_g)$ in LV polarization XAS is relatively smaller compared to LH, leading to the thickness modulations of electronic structure anisotropy in $RuO_2$ (110) films.

The $I(t_{2g})/I(e_g)$ at O $K$-edge XAS is proportional to the Ru-O orbital hybridization strength and the number of unoccupied Ru states (45). Since the Ru $L_3$-edge XAS spectrum in Fig. S4 shows the robust + 4 oxidation state for our $RuO_2$ (110) films, the increase in $I(t_{2g})/I(e_g)$ with film thickness indicates enhancement of Ru-O hybridization with strain-relaxations. We note that the enhancements of $I(t_{2g})/I(e_g)$ in XAS are more pronounced compared to our DFT calculations, which might suggest the additional strain effect such as structural phase transition (32) and/or dimensional crossover. On the other hand, because the anisotropic strain relaxation effect is more pronounced in the strongly bonded $e_g$-$\sigma^*$ states compared to the $t_{2g}$-$\pi^*$ bonding states (40), this leads to a more significant peak shift for the $e_g$-$\sigma^*$ state, resulting in the enhanced $\Delta$. Thus, we attribute the increase of $\Delta$ and $I(t_{2g})/I(e_g)$ above 4 nm $RuO_2$ films originating from the strain-relaxation-induced evolution in the electronic structure. We could not observe a specific thickness-dependent bandwidth change for each XAS peak, which might be due to the peak broadness caused by photoelectron damping. Nevertheless, the combined experimental XAS data and DFT calculations show how thickness and anisotropic strain relaxation along [001] results in the anisotropic electronic structure of $RuO_2$ films.

Finally, we investigate in Fig. 3 the polarization-selective optical anisotropy and carrier dynamics in $RuO_2$ (110) films by measuring the transient reflectivity change $\Delta R$ in response to ultrashort laser pulses using a femtosecond laser. Figure 3A shows the $\Delta R$ in the picosecond time domain for epitaxial $RuO_2$ (110) films of varying thicknesses. Both pump and probing laser have a center wavelength of 785 nm (1.58 eV) as illustrated in Fig. 3C to selectively excite and monitor the anisotropic Ru $t_{2g}$-related transitions. The polarization of the pump laser was applied along [001] direction whereas the transient optical reflectivity as a function of time, the $\varDelta R$, was measured with two different polarization angles (θ) of probing laser pulse (θ = 0° for $E_{[1\ 1\ 0]}$ and θ = 90° for $E_{[001]}$) for $RuO_2$ films of thicknesses 4, 7, and 12 nm. Data from other film thicknesses (2 and 17 nm) are shown in Fig S5. Since the optical penetration depth of $RuO_2$ is estimated to be ~ 26 nm from ellipsometry results, we confirmed that the entire film thickness was probed. Interestingly, the $\varDelta R$ at this timescale depends only on the polarization of the probing laser but remains independent of the polarization of the pump laser (Fig. S6). This indicates that the observed anisotropic hot carrier dynamics are intrinsic to the material's light-matter interaction, rather than dictated by the initial excitation conditions. Therefore, we focus on the polarization dependence of the probing laser. Figure 3A reveals that the difference in $\varDelta R$ between the two probe polarizations ($E_{[1\ 1\ 0]}$ and $E_{[001]}$) becomes increasingly pronounced as anisotropic strain relaxation increasing thickness $d$. More specially, as $d$ increases, the decay slope of $\varDelta R$ for $E_{[001]}$, where the electrical field is along the anisotropic strain relaxation direction, decreases. In contrast, the $\Delta R$ of the bulk RuO$_2$ (110) single crystal exhibits a negligible anisotropic behavior along both in-plane [1 1 0] and [001] directions (Fig. S7). Fig. 3B shows the anisotropic factor $\rho$ as a function of time. As $d$ increases, $\rho$ increases for all time ($t$). Notably, the value of $\rho$ for $t \leq 5$ ps is larger than that for $t > 5$ ps, indicating variations in dominant carrier dynamics depending on the time scale. To examine this further, we plot in Fig. 3D the optical anisotropy $\rho$ as a function of $d$ for select $t$ of 0.1, 1.5, 5, and 12 ps. Regardless of different time scales, the value of $\rho$ remains nearly unchanged for $d \leq 4$ nm. Whereas it becomes much larger for a smaller time scale ($\leq 1.5$ ps) for $d > 4$ nm. The highest $\rho$ value of anisotropically strain-relaxed films ($d > 4$ nm) below 1 ps suggests a significant role of electron-electron scatterings in the non-equilibrium state in determining transient optical anisotropy. At 0.1 ps, electron-electron scattering dominates the hot carrier dynamics, and then electron-phonon scattering becomes crucial until 12 ps. In Fig. 3A of anisotropically strain-relaxed 12 nm film, $\Delta R$ of E$_{[001]}$ exhibits a more gradual rise in $\Delta R$ below 2 ps compared to that of E$_{[1\,1\,0]}$ resulting in a large difference of $\rho$ values. The slower electron-electron thermalization time for E$_{[001]}$ (but not for E$_{[1\,1\,0]}$) indicates an anisotropic reduction of electron-electron scattering along [001] direction.

We further fit the anisotropic $\Delta R$ using two exponential components, A and B, shown in Fig. 3A. Since the fast decay component A in E$_{[001]}$ above 12 nm disappears (Fig. S5), we only included the slow decay component B in those cases. Fig. 3E shows thickness-dependent decay time for the A component ($\tau_A$). In strained films below 4 nm, $\tau_A$ value of E$_{[001]}$ is much smaller than that of E$_{[1\,1\,0]}$, and it increases and disappears above 12 nm as thickness increases. On the other hand, $\tau_A$ value of E$_{[1\,1\,0]}$ starts to decrease above 4 nm, and it finally becomes much faster than that of E$_{[001]}$. This opposing tendency between E$_{[001]}$ and E$_{[1\,1\,0]}$ induces and increases the anisotropy in the carrier dynamics with increasing film thickness. The evolutions of anisotropy with thickness are also confirmed in electrical transports using the two Hall bar devices along the [001] and [1 1 0] directions of RuO$_2$ (110) films, as shown in Fig. S8 (*46*). Figs. 3F and 3G show more detailed θ-resolved $\Delta R$ contour plots with fitting results of $\tau_A$ for 4 and 12 nm RuO$_2$ film, respectively. Both $\Delta R(\theta)$ contour plot and $\tau_A(\theta)$ of 12 nm show strong two-fold anisotropy, whereas those of 4 nm exhibit weak anisotropy. Fig. S9 further exhibits the same two-fold anisotropy for the 17 nm case. More specifically for 12 nm film, when θ is close to 90° or 270° parallel to the anisotropically strain-relaxed [001] crystal axis, $\tau_A$ becomes larger and merges to the slow B component at 90° or 270°. Interestingly, in contrast to the appearance of transient anisotropy at cryogenic temperature for Sr$_2$RuO$_4$ (*12*), the anisotropy between E$_{[001]}$ and E$_{[1\,1\,0]}$ of 4 and 12 nm RuO$_2$ films increases with anisotropic enhancement of $\Delta R$ for E$_{[1\,1\,0]}$ as temperature increases (Figs. S10 and S11). This further has a practical implication for robust optical anisotropy over a wide range of temperatures.

The anisotropic transient behavior in $\Delta R$ can originate from two possible mechanisms: instantaneous anisotropic carrier distribution or anisotropic carrier transitions. The former arises from an uneven distribution of excited carriers across momentum states (*23, 25, 27, 47*), while the latter is governed by direction-dependent optical transitions by the non-equilibrium population influenced by band structure asymmetry (*26, 27*). Both effects have been observed in pump-probe measurements of anisotropic materials (*23, 25-27, 47*). However, the insensitivity of $\Delta R$ to the pump laser polarization effectively rules out the contribution of an anisotropic carrier distribution, indicating that the transient carriers have already reached a quasi-equilibrium state with an elevated electronic temperature at picoseconds timescale. This suggests that the observed anisotropy in $\varDelta R$ stems from instantaneous anisotropic optical transitions along the [1 1 0] and [001] directions.

To investigate the origin of optical transient anisotropy, we first examine the anisotropic optical spectra of $RuO_2$ films using spectroscopic ellipsometry. Figs. 4A and 4B show optical spectra of strain-relaxed 17 nm $RuO_2$ thin films (see Supplementary Texts S1 and S2, Figs. S12 and S13 for further details). Here, $\sigma_{1, o}(\omega)$ and $\sigma_{1, e}(\omega)$ are real parts of ordinary (along [1 1 0] direction) and extraordinary (along [001] direction) optical conductivity spectra, respectively. Using Drude–Lorentz analysis, we extracted the individual spectra weight for four optical transitions (Lorentzian oscillators assigned by α, β, A, and γ) and free charge carrier contribution (Drude model). At the transition energy of 1.58 eV (i.e. at 785 nm of the pump/probe laser wavelength), both the Drude contribution and the Ru-$4d$ $t_{2g}$ → $t_{2g}$ interband transitions (α and β transitions) contribute to the optical response. Figs. 4C and 4D exhibit distinct thickness dependence of $\sigma_{1, o}(\omega)$ and $\sigma_{1, e}(\omega)$ due to anisotropic strain relaxations, respectively. As thickness increases, spectra weight of $\sigma_{1, e}(\omega)$ above 1.5 eV significantly increases compared to that of $\sigma_{1, o}(\omega)$, indicating the optical anisotropy in optical transitions for anisotropically strain-relaxed $RuO_2$. These results highlight the interplay between free carrier dynamics and interband transitions in governing the observed anisotropic optical response in strain-relaxed $RuO_2$.

The comparatively stronger interband contribution along the [001] direction arises from band nesting and orbital hybridization, which preferentially enhance transition matrix elements along this axis. Under photoexcitation, transient increase in carrier densities within the few-picosecond regime is modeled in our DFT calculations by shifting the chemical potentials ($\delta$), $\mu = \mu_0 + \delta$, where $\mu$ and $\mu_0$ correspond to the chemical potentials of the excited and ground states, respectively. The corresponding theoretical evaluations of $\sigma_{1, o}(\omega)$ and $\sigma_{1, e}(\omega)$ are presented in Figs. 4E-4H, showing the calculated interband contributions to $\sigma_{1, o}(\omega)$ and $\sigma_{1, e}(\omega)$ for anisotropically strain-relaxed (Figs. 4E and 4F) and fully strained $RuO_2$ (Figs. 4G and 4H). The transient state is captured by varying $\delta$ from 0 to 0.3 eV. In the ground state ($\delta = 0$), two prominent peaks appear near 0.7 and 1.2 eV, originating from Ru-$4d$ $t_{2g}$ → $t_{2g}$ transitions which is consistent with the experimental observation of α peak at ~ 0.7 eV. Our calculations also captured the effect of anisotropic strain relaxation: the peak intensities of the $t_{2g}$ → $t_{2g}$ transitions for $\sigma_{1, e}(\omega)$ are significantly stronger compared to those for $\sigma_{1, o}(\omega)$ in anisotropically strain-relaxed (Figs. 4E and 4F), which is in agreement with the experimental results (peak intensity corresponding to α). In the excited state ($\delta > 0$), as $\delta$ increases, $\sigma_{1, e}(\omega)$ of anisotropically strain-relaxed $RuO_2$ shows relatively small variations, particularly above 1.3 eV (Fig. 4F), while that of fully strained $RuO_2$ exhibits noticeable changes in the same energy window (Fig. 4H). The $\sigma_{1, o}(\omega)$ of both strain-relaxed and strained $RuO_2$ increases with $\delta$ (Figs. 4E and 4G). We note that the lack of dependence of $\sigma_{1, e}(\omega)$ for strain-relaxed $RuO_2$ and the slow decay of hot carriers along [001] for 12 nm film (Fig 3A) align with the transient $\Delta R$ observed in Fig. 3.

To understand this observation, we further examine the electronic band structures of fully strained and anisotropically strain-relaxed $RuO_2$, as shown in Fig. 5, depicting the band structure in the $k_x - k_y$ plane at $k_z = 0$ for intuitive visualization. For strain-relaxed $RuO_2$ in Fig. 5A, we observed two sets of nested bands along $\Gamma - X$ line (the zone center to [001] direction), with energy differences ($E_C - E_V$) between the nested bands estimated to be 0.7 eV ($V_1$ and $C_1$, green lines) and 1.2 eV ($V_2$ and $C_2$, blue lines), corresponding to peak positions of Ru-$4d$ $t_{2g}$ → $t_{2g}$ transitions in Figs. 4F and 4H. The existence of the nested bands along [001] direction suggests van Hove singularities in their joint density of states. It also explains why optical transitions are immune to changes in δ, consistent with the δ-independence of $\sigma_{1,e}(\omega)$. To further investigate the degree of band nesting, we visualize the $C_2$ and $V_2$ band structures away from the high symmetry lines (Figs. 5C and 5D). The contour plots in the bottom plane of Figs. 5C and 5D show the energy difference between the $C_2$ and $V_2$ bands ($E_{C2} - E_{V2}$) projected in the two-dimensional $k$ space. For anisotropically strain-relaxed $RuO_2$ (Fig. 5C), robust band nesting i.e., identical $E_{C2} - E_{V2}$ values, also occurs besides the Γ – X line, indicating nearly δ-independent $\sigma_{1,e}(\omega)$. On the other hand, for fully-strained $RuO_2$ (Fig. 5B and 5D), strain-induced lifting of degeneracy results in the splitting into $C_2$ and $C'_2$, reducing the degree of band nesting (decreasing $E_{C2} - E_{V2}$ values near X). This effect leads to a relatively pronounced δ-dependence of $\sigma_{1,e}(\omega)$ (Fig. 5D). We also note that the anisotropic δ-independence of $\sigma_{1,e}(\omega)$ aligns with the temperature-independent $\Delta R$ and $\tau_A$ for $E_{[001]}$ exclusively in the anisotropically strain-relaxed $RuO_2$ (Fig. S12D). These results indicate that anisotropic strain relaxation effectively modifies band nesting, providing a control mechanism for tuning transient optical responses in metallic epitaxial $RuO_2$ films.

## Summary

In summary, we realized an unexpected yet systematic modulation of electronic anisotropy in metallic $RuO_2$ heterostructures by epitaxial strain control. DFT calculations revealed the anisotropic strain-dependent electronic structure with an enhancement of orbital hybridizations due to the anisotropic strain relaxation along the [001] direction. Following the theoretical insights of DFT, comprehensive observations of XAS, spectroscopic ellipsometry, and optical pump-probe techniques for $RuO_2$/$TiO_2$ (110) films confirmed the large enhancement of anisotropy in the electronic structure and hot carrier dynamics in anisotropically strain-relaxed $RuO_2$ (110) films. We note that the critical thickness for anisotropic strain relaxation deduced from the optical spectroscopies is consistent with abrupt changes in the symmetry patterns of the optical SHG above 4 nm of $RuO_2$/$TiO_2$ (110) films (*32*). The consistency across independent experiments suggests that strain relaxation begins around 4 nm and dominates the $RuO_2$ film thickness dependence of various properties. Our epitaxy approach offers a facile method for realizing ultrafast anisotropic electronic dynamics of metallic rutile systems, which is distinct from conventional studies in insulating/semimetal systems (*23, 25-27*). Furthermore, our findings establish a fundamental link between band nesting and transient optical anisotropy, shedding light on the underlying mechanisms governing directional carrier dynamics in metals. This work not only advances the fundamental understanding of strain-driven optical anisotropy in high-symmetry metals but also opens new pathways for the design of next-generation optoelectronic devices with precise tunability via epitaxial strain control.

## Methods

### Hybrid MBE and structural characterizations

Epitaxial $RuO_2$ films were grown using an oxide hMBE system (Scienta Omicron) on $TiO_2$ (110) single crystalline substrates (Crystec). The substrate treatment process involved cleaning with acetone, methanol, and isopropanol, followed by a 2-hour bake at 200 °C in a load lock chamber. Before film growth, a 20-minute annealing in oxygen plasma at a growth temperature of 300 °C was performed. A metal-organic precursor, Ru(acac)$_3$, was thermally evaporated using a low-temperature effusion cell (MBE Komponenten) with an effusion cell temperature of between 170 to 180 °C. The radio frequency oxygen plasma and an oxygen gas pressure of $5 \times 10^{-6}$ Torr were used during the growth. After the growth, the sample was cooled to 120 °C in the presence of oxygen plasma to prevent the potential formation of oxygen vacancies. We monitored film surfaces before, during, and after growth using in-situ reflection high-energy electron diffraction (Staib Instruments). Additionally, we determined the crystallinity, film thickness, roughness, and strain state using XRD (Rigaku SmartLab XE) with reciprocal space mapping (RSM), X-ray reflectivity, and θ-2θ measurements. Surface morphologies were measured using atomic force microscopy (Bruker Nanoscope V Multimode 8) with peakforce tapping mode. In this study, we focused on samples with film thicknesses below 20 nm for several reasons: (1) to ensure that the pump-probe experiments remained within the probing depth ~26 nm; (2) to avoid surface cracking that occurs in films thicker than 20 nm; and (3) to isolate anisotropic strain relaxation along [001] because we have confirmed strain relaxation also occurs along the [1 1 0] direction at 26 nm in the previous study (*35*).

**Single crystal synthesis**

RuO$_2$ (001) and (110) single crystals were synthesized by sintering a pelletized RuO$_2$ powder inside an alumina crucible in the air. The pellet was heated up to 1473 K and slowly cooled down to 1023 K at a cooling rate of 1.25 K per hour until the furnace was turned off for quenching. The blue-black metallic crystals were at maximum $1 \times 1 \times 0.5$ mm$^3$ in size.

**Density functional theory calculation**

We performed first-principles calculations based on density functional theory (*48*) as implemented in the Vienna *ab initio* simulation package (*49*). The projector augmented wave potentials (*50, 51*) were used to describe the valence electrons and the plane-wave kinetic energy cutoff was chosen to be 500 eV. The exchange-correlation function was treated by the generalized gradient approximation (GGA) of Perdew-Burke-Ernzerhof (*52*). The lattice parameters are summarized in Table S1. The corresponding Brillouin zones were sampled with a $20 \times 10 \times 10$ *k*-grid mesh. When projecting the density of states, the following notations were used for the coordinate axes: x = [001], y = [1 1 0], and z = [110]. The optical conductivity was calculated by the Kubo-Greenwood formula defined as (*53, 54*),

$$\sigma_{\alpha\beta}(\hbar\omega) = \frac{ie^2\hbar}{N_k\Omega_C} \sum_{\mathbf{k}} \sum_{n,m} \frac{f_{m\mathbf{k}} - f_{n\mathbf{k}}}{\varepsilon_{m\mathbf{k}} - \varepsilon_{n\mathbf{k}}} \frac{\langle\psi_{n\mathbf{k}}|v_\alpha|\psi_{m\mathbf{k}}\rangle\langle\psi_{m\mathbf{k}}|v_\beta|\psi_{n\mathbf{k}}\rangle}{\varepsilon_{m\mathbf{k}} - \varepsilon_{n\mathbf{k}} - (\hbar\omega + i\eta)},$$

where $\alpha, \beta$ are Carteisan directions, $N_k$ is the total number of k-points, and $\Omega_C$ is a volume of the unit cell structures. Here, a broadening parameter of $\eta$ is chosen to be 100 meV. The numerical calculations of $\sigma_{\alpha\beta}(\hbar\omega)$ were performed by the Wannier90 package (*55*). To construct the Wannier Hamiltonian, we used *d* and *p* orbital projections for Ru and O atoms, respectively.

**X-ray absorption spectroscopy**

O *K*-edge XAS spectra of RuO$_2$/TiO$_2$ thin films are measured at the 6A beamline of the Pohang Light Source using TEY measurement, which is sensitive to the film surface. The measurements were conducted at room temperature and surface normal configuration with π (σ) polarization, where the electrical field direction of the X-rays is parallel to [001] ([1 1 0]) crystal direction of $RuO_2/TiO_2$ (110).

### Spectroscopic ellipsometry

To acquire the optical conductivity of $RuO_2/TiO_2$ (110) heterostructures, we employed spectroscopic ellipsometry (M-2000, J. A. Woollam Co., Inc.) with a wavelength range of 0.73 to 6.44 eV and incident angles of 60°, 65°, and 70°. To experimentally access the anisotropic optical response of rutile (110) systems using spectroscopic ellipsometry, we measured each sample twice by rotating the crystal orientation by 90°: 1) *p*-polarization along [001] and *s*-polarization along [1 1 0] 2) *p*-polarization along [1 1 0] and *s*-polarization along [00 1 ]. The detail of the anisotropic model is included in Supplementary Text S1. This approach yielded consistent results with previously reported optical constants for bulk $TiO_2$ (data not shown). Subsequently, we analyzed the thin film samples using the same method and meticulously evaluated the dielectric function of the $RuO_2$ thin film using a uniaxial layer model.

### Optical pump-probe experiment

We measured changes in transient reflectivity ($\Delta R$) using the optical pump-probe method to monitor the anisotropic carrier dynamics. The pump and probe beam were separated by a beam-splitter from an 80 MHz pulsed laser beam that has 785 nm center wavelengths (Vision-S, Coherent). We modulated the pump and probe beam using an electro-optic modulator with 10 MHz and a mechanical chopper with 700 Hz, respectively. Both pump and probe beam were focused on the samples in the normal incidence, using a 5x microscope lens with a beam size of 20 μm. The power of the pump and probe beam was set to be 50 mW and 3 mW, respectively. To block the pump beam from entering the detector, we employed a two-tint method using the sharp-edge long-pass filter and short-pass filter (Semrock). By using a two-channel digital Lock-in amplifier (Zurich instrument), we demodulated the signal with side-band frequencies composed of 10 MHz and 700 Hz to achieve the pump-induced reflectivity changes. To focus on the decay of hot carrier dynamics, we extracted the oscillation component from the coherent phonon oscillation in Fig. S14.

### Electrical transport using Hall bar devices

We fabricated the Hall bar devices along two different in-plane [001] and [110] directions utilizing photolithography followed by argon ion milling (*46*). We used Aluminum (Al) wire bonding for electrical contacts. The electrical conductivities (σ) are measured by using Quantum Design Dynacool Physical Property Measurement System (PPMS) with Keithley 2430 source-measure unit.

**Acknowledgments**

**Funding:** Film synthesis (S.G.J and B.J.) was supported by the U.S. Department of Energy through grant Nos. DE-SC0020211, and (partly) DE-SC0024710. Structural characterization, transport, and ellipsometry (at UMN) were supported by the Air Force Office of Scientific Research (AFOSR) through Grant Nos. FA9550-21-1-0025, and FA9550-24-1-0169. S.N. was supported partially by the UMN MRSEC program under Award No. DMR-2011401. Parts of this work were carried out at the Characterization Facility, University of Minnesota, which receives partial support from the NSF through the MRSEC program under Award No. DMR-2011401. The work by J.H.L. and C.K. was supported by the Global Research Development Center (GRDC) Cooperative Hub Program through the National Research Foundation of Korea (NRF) funded by the Ministry of Science and ICT(MSIT) (Grant No. RS-2023-00258359) and the NRF grant funded by MSIT (Grant No. NRF-2022R1A3B1077234). This work was supported by the National Research Foundation of Korea (NRF) grant funded by the Korea government (MSIT) (No. 2022R1A2C2007847 (I.H.C. and J.S.L.), 2021R1A2C2011340 (O.J. and W.S.C.), RS-2023-00220471 (O.J. and W.S.C.), and RS-2023-00281671(O.J. and W.S.C.)).

**Author contributions:**
    Conceptualization: S.G.J, I.H.C., S.L., J.S.L., T.L. B.J.
    Methodology: S.G.J, I.H.C., S.L., J.Y.O., S.N., J.H.L., C.K., A.S., W.S.C., T.L. J.S.L., B.J.
    Visualization: S.G.J, I.H.C.
    Supervision: J.S.L., B.J.
    Writing—original draft: S.G.J, I.H.C., S.L., J.S.L., T.L. B.J.
    Writing—review & editing: S.G.J, I.H.C., S.L., J.Y.O., S.N., J.H.L., C.K., A.S., W.S.C., T.L. J.S.L., B.J.

**Competing interests:** All other authors declare they have no competing interests.

**Data and materials availability:** All data are available in the main text or the supplementary materials.

# Figures

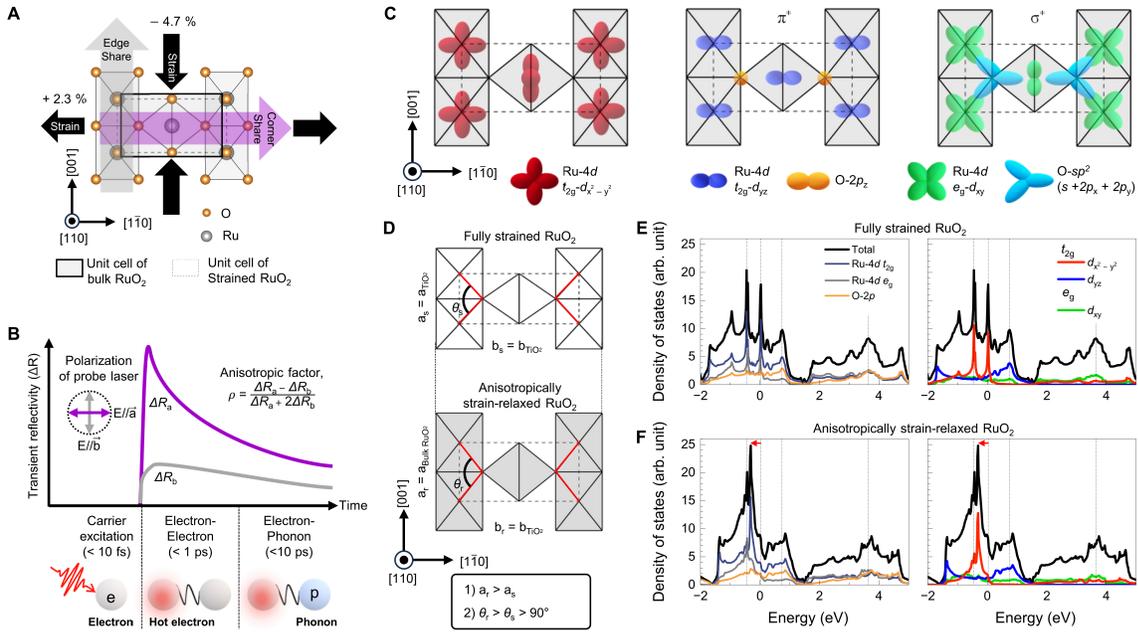

**Fig. 1. Epitaxial strain control of electronic anisotropy in RuO₂/TiO₂ (110) heterostructure.** (**A**) Schematic illustrates anisotropic ionic bonding in RuO₂ (110) heterostructure with anisotropic epitaxial strain (black arrows) on TiO₂ (110) substrate. The strain values are estimated based on bulk lattice parameters. (**B**) Schematic illustration of *ΔR* as a function of time delay using an optical pump-probe demonstrates an experimental example of transient electronic anisotropy. The bottom panels show the dominant carrier dynamics exited by pumping laser with different ultrafast time scales. (**C**) Schematic representations of three representative orbital states projected on the RuO₂ (110) plane. (**D**) Schematics of the RuO₂ model systems are adopted for DFT calculations to describe anisotropic strain relaxations along [001] direction. The anisotropic strain-relaxation yields two structural evolutions: 1) $a_r > a_s$ and 2) $\theta_r > \theta_s > 90°$, as discussed in the Main Text. (**E** and **F**) The density of states projected by Ru-4*d* and O-2*p* states (left panels) and detailed Ru-4*d* states ($d_{x^2-y^2}$, $d_{yz}$, and $d_{xy}$, right panels) for (**E**) fully strained and (**F**) anisotropically strain-relaxed RuO₂. The vertical dotted lines are each peak position of the orbital states for the fully strained case. The red arrows indicate the [001] strain relaxation-induced energy shift of $d_{x^2-y^2}$ state.

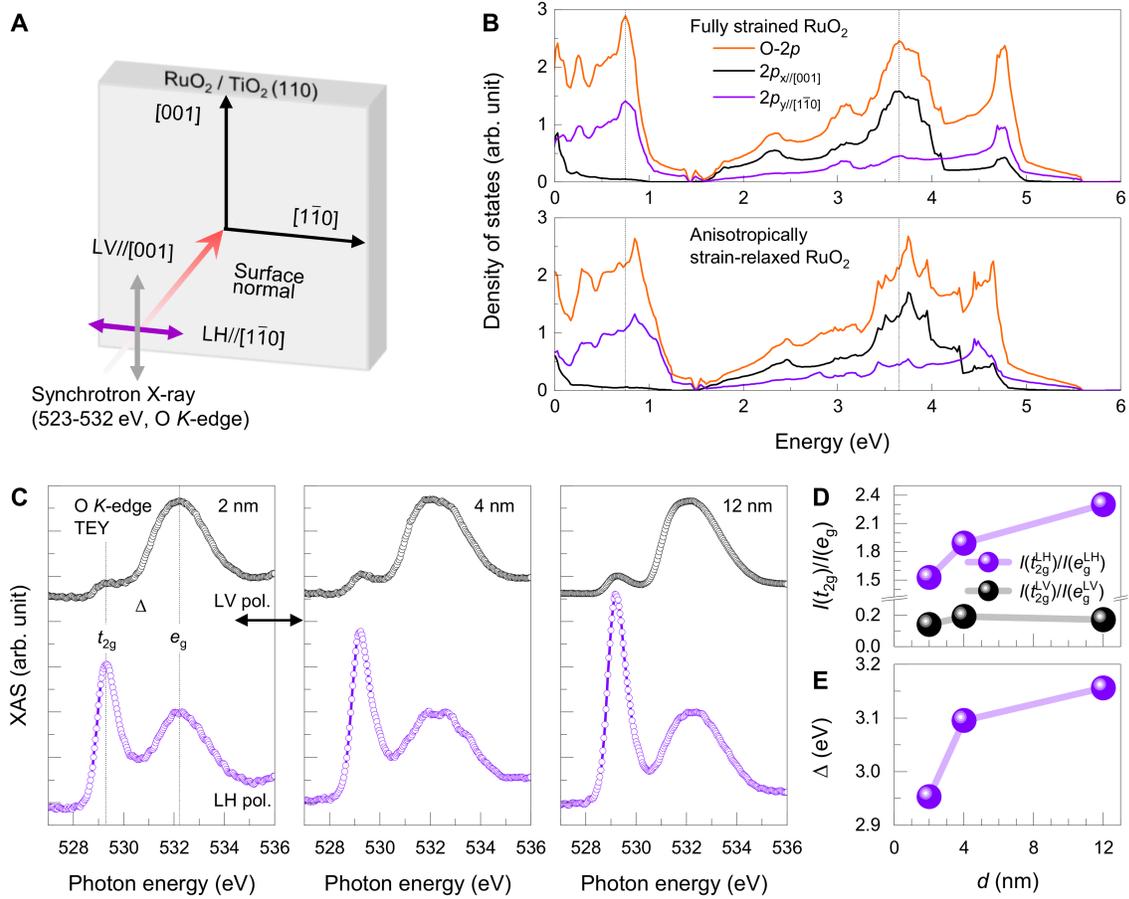

**Fig. 2. Modulated anisotropic Ru-O hybridizations of epitaxial RuO$_2$/TiO$_2$ (110) heterostructures.** (**A**) Schematic of the scattering geometry with the surface normal incident synchrotron X-ray beam. LV and LH polarizations are parallel to [001] and [1 1 0] directions of RuO$_2$/TiO$_2$ (110) samples, respectively. (**B**) The density of states for O-2$p$ orbitals are projected to [001] (2$p_x$) and [1 1 0] (2$p_y$) direction for fully strained (lower panel) and anisotropically strain-relaxed RuO$_2$ films (upper panel). (**C**) XAS spectra at the O $K$-edge measured by TEY methods show LH (purple) and LV (black) polarization dependence for RuO$_2$ thicknesses of 2, 4, and 12 nm in RuO$_2$/TiO$_2$ (110). Vertical dotted lines represent the peak positions indicating Ru-4$d$ $t_{2g}$ and $e_g$ with O-2$p$ hybridizations with their energy difference, Δ, for the 2 nm case. (**D**) $I(t_{2g})/I(e_g)$ and (**E**) Δ shows RuO$_2$ thickness dependence due to the anisotropic strain relaxations. The $I(t_{2g})/I(e_g)$ was assigned based on the point of highest intensity for each $t_{2g}$ and $e_g$ related peak.

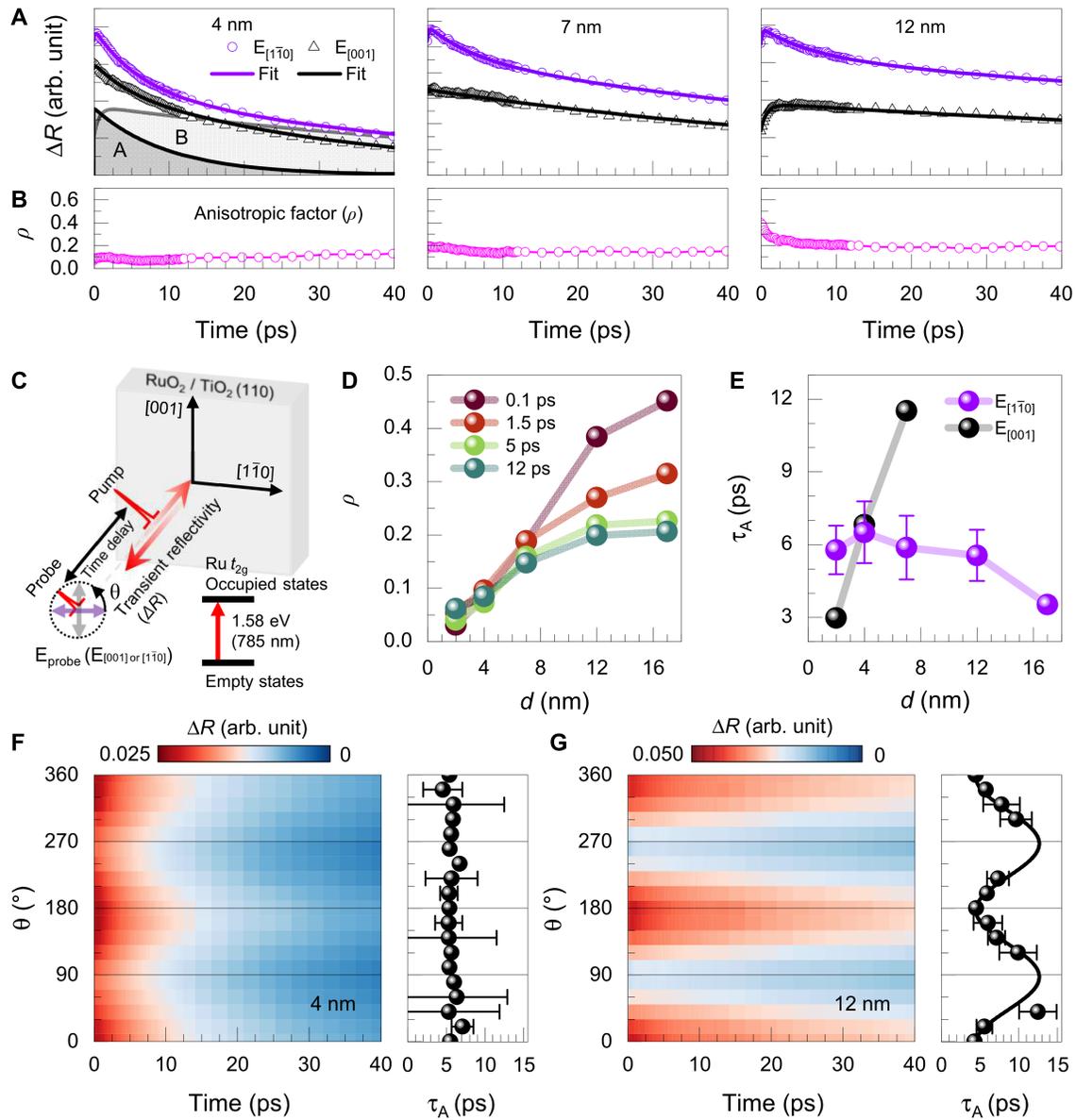

**Fig. 3. Anisotropic ultrafast carrier dynamics of epitaxial RuO$_2$/TiO$_2$ (110) heterostructures.** (**A**) Transient reflectivity change ($\Delta R$) of RuO$_2$/TiO$_2$ (110) films for two perpendicular polarizations of probe laser along [1 1 0] and [001] of crystal axis with (**B**) anisotropic factor, $\rho$. (**C**) Schematic of the scattering geometry with surface normal incident optical pump-probe measurement. (**D**) Time-dependent $\rho$ and (**E**) polarization-dependent anisotropic $\tau_A$ with different film thickness ($d$). Polarization angle (θ) resolved $\Delta R$ (left panel) with $\tau_A$ (right panel) for (**F**) 4 and (**G**) 12 nm RuO$_2$ cases. The solid line of $\tau_A(\theta)$ represents the guideline for easy to eye using the sine function.

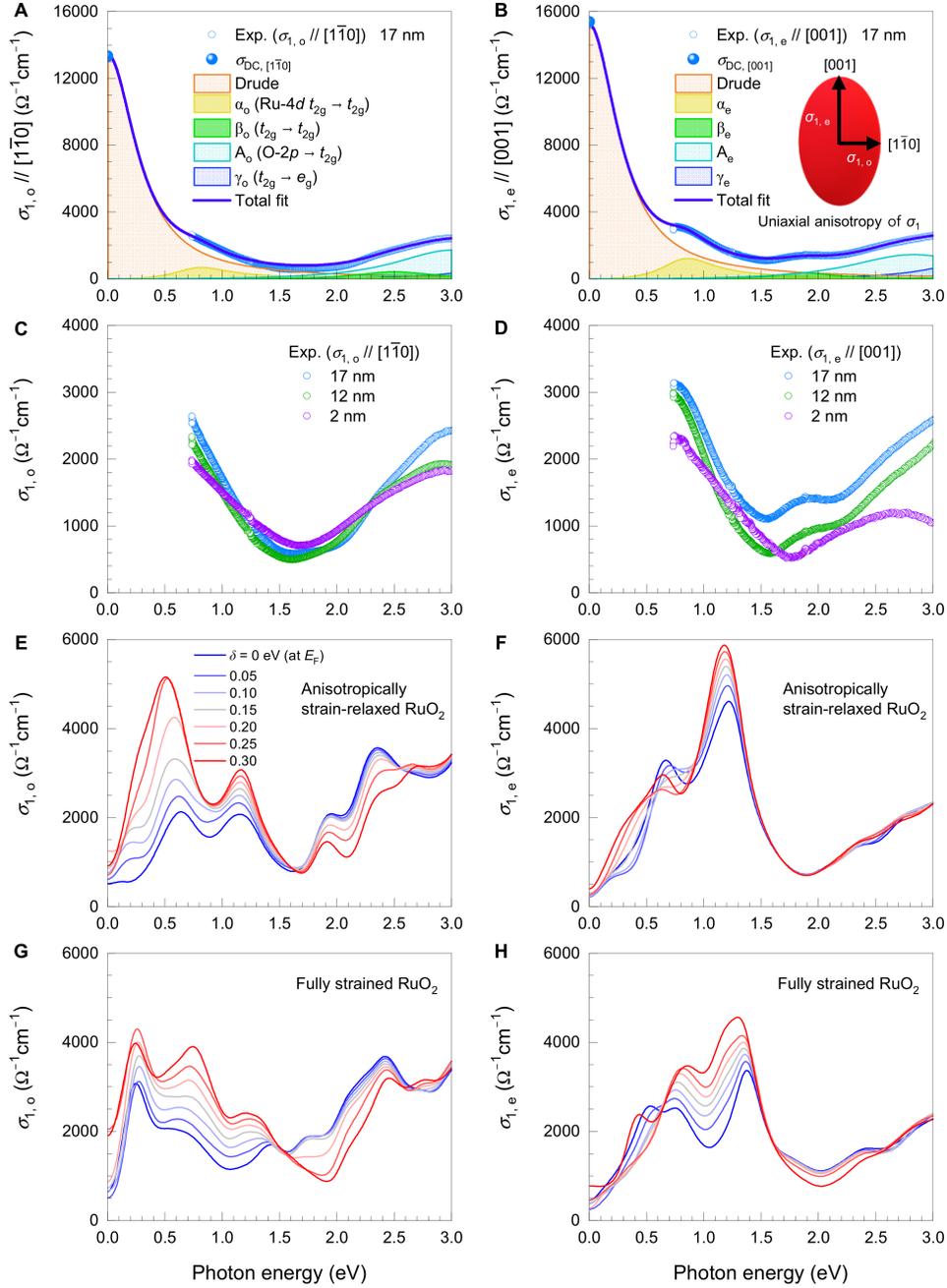

**Fig. 4. Anisotropic optical conductivity spectra from experiment and DFT calculations.**
(**A**) $\sigma_{1,o}(\omega)$ along [1 1̄ 0] direction and (**B**) $\sigma_{1,e}(\omega)$ along [001] direction of anisotropically strain-relaxed 17 nm RuO$_2$ (110) film. The shaded area indicates the estimated optical transitions including a single Drude model and Lorentzian oscillators. Schematic in (**B**) shows the uniaxial anisotropic model adopted in optical spectrum analysis. (**C**) $\sigma_{1,o}(\omega)$ and (**D**) $\sigma_{1,e}(\omega)$ show anisotropic thickness-dependence related to the anisotropic strain relaxation along [001] direction in RuO$_2$/TiO$_2$ (110). Interband contributions of $\sigma_{1,e}(\omega)$ and $\sigma_{1,o}(\omega)$ for (**E** and **F**) anisotropically strain-relaxed RuO$_2$ and (**G** and **H**) fully strained RuO$_2$ calculated by DFT with a modified chemical potential ($\mu = \mu_0 + \delta$).

Page 17 of 18

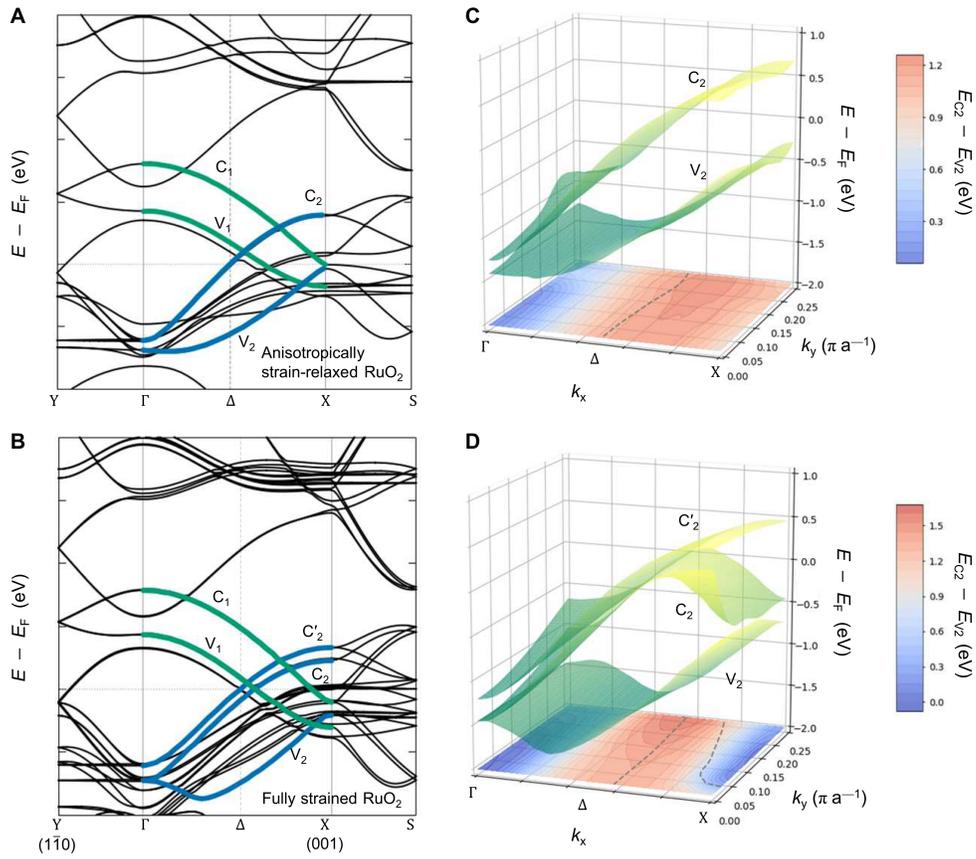

**Fig. 5. Strain-controlled electronic band nesting.** Electronic structures along the in-plane high-symmetry line and the nested bands on the two-dimensional $k$ mesh for (**A** and **C**) anisotropically strain-relaxed $RuO_2$ and (**B** and **D**) fully strained $RuO_2$. The color contour maps in the bottom plane show the $E_{C2} - E_{V2}$ values and the grey dashed line indicates the zero-energy level of $C_2$ band, indicating $k$ space allowing optical transition ($C_2 > 0$ and $V_2 < 0$).